# Using the Modified Allan Variance for Accurate Estimation of the Hurst Parameter of Long-Range Dependent Traffic

Stefano Bregni*, *Senior Member, IEEE,*
Luca Primerano

Politecnico di Milano, Dept. of Electronics and Information
Piazza Leonardo Da Vinci 32, 20133 Milano, Italy.
Tel.: +39-02-2399.3503. Fax: +39-02-2399.3413
E-mail: bregni@elet.polimi.it (*managing author)

## ABSTRACT

Internet traffic exhibits self-similarity and long-range dependence (LRD) on various time scales. A well studied issue is the estimation of statistical parameters characterizing traffic self-similarity and LRD, such as the Hurst parameter H. In this paper, we propose to adapt the Modified Allan Variance (MAVAR), a time-domain quantity originally conceived to discriminate fractional noise in frequency stability measurement, to estimate the Hurst parameter of LRD traffic traces and, more generally, to identify fractional noise components in network traffic. This novel method is validated by comparison to one of the best techniques for analyzing self-similar and LRD traffic: the logscale diagram based on wavelet analysis. Both methods are applied to pseudo-random LRD data series, generated with assigned values of H. The superior spectral sensitivity of MAVAR achieves outstanding accuracy in *estimating* H, even better than the logscale method. The behaviour of MAVAR with most common deterministic signals that yield nonstationarity in data under analysis is also studied. Finally, both techniques are applied to a real IP traffic trace, providing a sound example of the usefulness of MAVAR also in traffic characterization, to complement other established techniques as the logscale method.

**Note**

This paper is based in part on ideas presented in the preliminary version "The Modified Allan Variance as Time-Domain Analysis Tool for Estimating the Hurst Parameter of Long-Range Dependent Traffic", by S. Bregni and L. Primerano, appeared in *Proc. of IEEE GLOBECOM 2004*, Dallas, TX, USA, Dec. 2004.

This work was supported in part by the Italian Ministry of Education, University and Research (MIUR) under the research FIRB project TANGO.

# I. INTRODUCTION

Experimental measurements have revealed that Internet traffic exhibits high temporal complexity. Far from being dominated by well-identifiable pseudo-periodic components, Internet traffic features intriguing temporal correlation properties, such as self-similarity and long memory (long-range dependence) on various time scales [1]—[3]. Contrary to the classical Poisson-model assumption, these properties emphasize long-range time-correlation between packet arrivals. Fractional noise and fractional Brownian motion models are used to describe the behaviour of Internet traffic traces, which include, but are not limited to, cumulative or incremental data count transmitted over time, time series of successive TCP connection durations, inter-arrival time series of successive TCP connections or IP packets, etc.

In a self-similar random process, a dilated portion of a realization (sample path) has the same statistical characterization than the whole. "Dilating" is applied on both amplitude and time axes of the sample path, according to a scaling parameter called Hurst parameter. On the other hand, long-range dependence (LRD) is a long-memory property observed on large time scales: LRD is usually equated with an asymptotic power-law decrease of the autocovariance function and of the power spectral density (PSD). Under some hypotheses, the integral of a LRD process is self-similar (e.g., fractional Brownian motion, integral of fractional Gaussian noise).

The issue of estimating statistical parameters that characterize self-similar and LRD random processes has been often studied, aiming at best modelling of traffic for example to the purpose of network simulation. Among such quantities, the Hurst parameter $H$ has been devoted particular attention in literature. Several algorithms are used to estimate $H$ under various hypotheses [1][2][4][5].

In time and frequency measurement theory, on the other hand, the Modified Allan Variance (MAVAR) is a well known time-domain quantity conceived for frequency stability characterization of precision oscillators [6]—[10]. This variance was originally designed to discriminate effectively noise types with power-law spectrum (e.g., white, flicker and random-walk noise), recognized very commonly in frequency sources. Moreover, international standards (ANSI, ETSI and ITU-T) specify stability requirements for telecommunications network synchronization also in terms of a quantity directly derived from MAVAR, namely the Time Variance (TVAR) [11].

In this paper, we propose to adapt MAVAR to estimate the Hurst parameter of LRD traffic traces and, more generally, to identify fractional noise components in network traffic. In our knowledge, this is the first time that MAVAR is applied in this context. This method is validated by comparison to one of the best and most widely adopted algorithms for analyzing self-similar and LRD traffic: the logscale diagram (LD) technique based on wavelet analysis [2][4]. To this purpose, both methods are applied to pseudo-random LRD data series, generated with assigned values of the *H* parameter. The behaviour of MAVAR with most common deterministic signals that yield nonstationarity in data under analysis is also studied. Finally, both techniques are applied to a real IP traffic trace, providing a sound example of the usefulness of MAVAR also in traffic characterization, to complement other established techniques as the logscale diagram method.

## II. Self-Similarity and Long-Range Dependence

A random process $X(t)$ (say, cumulative packet arrivals in the time interval $[0, t]$) , with $t \in \Re$, is said to be *self-similar*, with scaling parameter of self-similarity or Hurst parameter $H > 0$, if

$$X(t) \overset{d}{=} a^{-H} X(at) \tag{1}$$

for all $a > 0$, where $\overset{d}{=}$ denotes equality for all finite-dimensional distributions [1][2]. In other terms, the statistical description of the process $X(t)$ does not change by *scaling* simultaneously its amplitude by $a^{-H}$ and the time axis by $a$. Self-similar processes are not stationary by definition, since the moments of $X(t)$, provided they exist, behave as power laws of time:

$$E\left[ \left| X(t) \right|^q \right] = E\left[ \left| X(1) \right|^q \right] \cdot |t|^{qH} \tag{2.}$$

In practical applications, the class of self-similar (H-SS) processes is usually restricted to that of *self-similar processes with stationary increments* (or H-SSSI processes), which are "integral" of some stationary process. For example, consider the $\delta$-increment process of $X(t)$, defined as $Y_\delta(t) = X(t)$-$X(t-\delta)$ (say, packet arrivals in the last $\delta$ time units). For a *H*-SSSI process $X(t)$, $Y_\delta(t)$ is stationary and $0<H<1$ [2].

*Long-range dependence* (LRD) of a process is defined by an asymptotic power-law decrease of its autocovariance or equivalently PSD functions [1][2]. Let $Y(t)$, with $t \in \Re$, be a second-order stationary stochastic process. The process $Y(t)$ exhibits LRD if its autocovariance function follows asymptotically

$$R_Y(\delta) \sim c_1 |\delta|^{\gamma-1} \quad \text{for } \delta \to +\infty \,, \, 0 < \gamma < 1 \tag{3}$$

or, equivalently, its power spectral density (PSD) follows asymptotically

$$S_Y(f) \sim c_2 |f|^{-\gamma} \quad \text{for } f \to 0 \,, \, 0 < \gamma < 1 \tag{4}.$$

A process with such a PSD is also known as fractional noise. The integral of fractional Gaussian noise is fractional Brownian motion. It can be proved [2] that H-SSSI processes $X(t)$ with $1/2 < H < 1$ have long-range dependent increments $Y(t)$, with

$$\gamma = 2H - 1 \tag{5}.$$

Strictly speaking, the Hurst parameter characterizes self-similar processes, but it is frequently used to label also the long-range dependent increment processes of H-SSSI processes. In this paper, we follow this common custom with no ambiguity. Hence, the expression "Hurst parameter of a LRD process" (characterized by the parameter $\gamma$) denotes actually, by extension, the Hurst parameter $H = (\gamma+1)/2$ of its integral H-SSSI parent process. Consistently, in traffic measurement, the Hurst parameter of a LRD bit-per-time-unit data sequence actually characterizes the self-similarity of the parent cumulative bit count sequence.

Several techniques are in use for quantifying the degree of self-similarity or long-range dependence in a generic time series (e.g., traffic trace) supposed adhering to this model. In particular, most attention has been devoted to the estimation of the Hurst parameter $H$ and of the power-law spectrum exponent $\gamma$ [1][2][4][5].

In the time domain, the so-called *variance-time plot* method studies the variance $\sigma^2(m)$ of aggregated time series $Y^{(m)}(n)$, made of samples computed by averaging nonoverlapping windows of width $m$ of the original stationary increment sequence $Y(n)$, supposed LRD. By observing the decay of $\sigma^2(m)$ with the window width $m$, it is possible to infer the spectrum power law and thus $\gamma$ and $H$.

In the frequency domain, a simple *periodogram* log-log plot yields a straightforward estimation of $\gamma$ and $H$ from its slope, according to (4). Nevertheless, it should be noted that this technique is not well suited to analyze processes with spectrum (4), which gather most power for $f\rightarrow0$, being periodogram samples evenly spaced in frequency.

More recently, a breakthrough occurred when techniques based on wavelet analysis were introduced [4]. Due to their sensitivity to scaling phenomena over a fine range of scales, wavelets are well suited to detect self-similarity or even other more complex scaling behaviours, such as multifractality. Among wavelet-based techniques, the so-called *logscale diagram* (LD) is of outmost importance. It analyzes the data series under study over an interval of scales (octaves), ranging from 1 (finest detail) to a longest scale given by the series finite length. By observing the diagram slope, $H$ and $\gamma$ can be estimated.

## III. The Modified Allan Variance

This section introduces the Modified Allan Variance and briefly summarizes some of its properties most relevant to our aim, referring the interested reader to a comprehensive bibliography for all details. Moreover, its behaviour with generic power-law random noise and most common deterministic signals that yield nonstationarity in data under analysis is also studied.

## A. Background

In stability characterization of precision oscillators, common models for spectral densities of phase and frequency noise are power laws similar to (4). In experimental results of phase measurement, integer values $0 \leq \gamma \leq 4$ are commonly found (e.g., the phase follows a random walk, or Brownian motion, when instantaneous frequency is white noise). Although values $\gamma \geq 1$ yield model pathologies, such as infinite power (variance) and even nonstationarity[1], this model has been always widely adopted, considering also that real-world constraints imply measurement finite bandwidth and finite duration. For an excellent survey on characterization of phase and frequency instabilities, including model issues and mathematical tools, see Rutman [13].

To circumvent such pathologies, in particular the variance increasing indefinitely with the measurement interval if $\gamma \geq 1$, a useful approach is to evaluate the variance of the $M$-th derivative (supposed stationary) of the process. For example, the two-sample variance (a.k.a. Allan variance) recommended by IEEE in 1971 [14] for characterization of frequency stability in the time domain, following the pioneering work of D. W. Allan in 1966 [15], is a sort of variance of the second difference of phase measurement samples. The structure function theory, developed by Lindsey and Chie [16], gives a unifying view of time-domain quantities evaluated on the $M$-th increment (supposed stationary) of the random process under study. To probe further this wide subject, see also [6] and [17]—[20].

Among the various quantities defined in the time domain for stability characterization of precision oscillators, the Modified Allan Variance (MAVAR) Mod $\sigma_y^2(\tau)$ plays a prominent role [6]—[11]. This variance was conceived in 1981 by modifying the definition of the standard Allan variance, to improve its poor discrimination capability against white and flicker phase noise. Being based on the second difference of input data, it converges to finite values for all power-law noise types

---

[1]   In fractional Brownian motion, defining the precise meaning of "power spectral density", proportional to $|f|^{-\gamma}$ with $\gamma \geq 2$,

with $\gamma < 5$. Moreover, over the full range $0 \le \gamma < 5$, it allows accurate estimation of the parameter $\gamma$ and features numerous other advantages, as it will be shown in the following. These properties suggest its fruitful application also to LRD and self-similar network traffic analysis and call for a thorough investigation about its usefulness and limits in this field.

## B.  Definition in the Time Domain

Given an infinite sequence $\{x_k\}$ of samples of a signal $x(t)$, evenly spaced in time with sampling period $\tau_0$, the MAVAR is defined as

$$\text{Mod } \sigma_y^2(\tau) = \frac{1}{2n^2\tau_0^2} \left\langle \left[ \frac{1}{n} \sum_{j=1}^{n} \left( x_{j+2n} - 2x_{j+n} + x_j \right) \right]^2 \right\rangle \tag{6}$$

where the operator $<\cdot>$ denotes infinite-time averaging and $\tau = n\tau_0$ is the observation interval.

In time and frequency stability characterization, the data sequence $\{x_k\}$ is made of samples of random time deviation $x(t)$ of the chronosignal under test [6][13][14]. The average value of the random fractional frequency $y(t) = x'(t)$, measured over the observation interval $\tau$ beginning at the generic instant $t_k$, is the sample $y_k$ of the first difference of the sequence $\{x_k\}$ evaluated over $\tau$, i.e.

$$y_k(\tau) = \frac{1}{\tau} \int_{t_k}^{t_k+\tau} y(t)dt = \frac{x_{k+n} - x_k}{n\tau_0} \tag{7}.$$

Hence, the definition of MAVAR given in terms of the fractional frequency sequence $\{y_k\}$ is

$$\text{Mod } \sigma_y^2(\tau) = \frac{1}{2} \left\langle \left[ \frac{1}{n} \sum_{j=1}^{n} \left( y_{j+n} - y_j \right) \right]^2 \right\rangle \tag{8}.$$

---

raises some conceptual concern due to the nonstationarity of these processes. See [12] for a discussion and clarification of this point.

The MAVAR is thus a kind of variance of the first difference of the fractional frequency $y(t)$ or of the second difference of the time deviation $x(t)$. In very brief, it differs from the basic Allan variance in the additional average over $n$ adjacent measurements: for $n=1$ ($\tau=\tau_0$), the two variances coincide. In fact, the basic Allan variance is the plain time-average of half the square difference of couples of frequency samples (cf. eq. (8)). For more details on MAVAR rationale and properties, see [6]—[10].

It is immediate to translate these concepts to traffic analysis, letting the frequency process $y(t)$ be the packet or bit rate and the time deviation $x(t)$ be the cumulative packet or bit count since $t=0$. Nevertheless, nothing prevents from evaluating (6) on a sequence $\{x_k\}$ made of packet or bit rate samples, for example if a rate drift is suspected in these data, similarly to increasing the number of vanishing moments in logscale diagram computation.

In practical measurements, given a finite set of $N$ samples $\{x_k\}$, again spaced by the sampling period $\tau_0$, an estimate of MAVAR can be computed using the ITU-T standard estimator [6][11]

$$\text{Mod}\,\sigma_y^2(n\tau_0) = \frac{1}{2n^4\tau_0^2(N-3n+1)}\sum_{j=1}^{N-3n+1}\left[\sum_{i=j}^{n+j-1}\left(x_{i+2n}-2x_{i+n}+x_i\right)\right]^2 \tag{9}$$

with $n=1$, 2, ..., $\lfloor N/3 \rfloor$. A recursive algorithm for fast computation of this estimator exists [6], which cuts down the number of operations needed to evaluate MAVAR, for all possible values of $n$, to $\sim N^2$ instead of $\sim N^3$.

It should be noted that the point estimate (9) of $\text{Mod}\,\sigma_y^2(\tau)$, computed by averaging $N$-$3n$+1 terms, is a random variable itself. Its variance can be computed and used to assess confidence interval estimates under several simplifying assumptions [21]—[26]. Exact computation of confidence intervals is not immediate and, annoyingly enough, depends on the spectrum of the underlying noise. However, in general, along a plot of $\text{Mod}\,\sigma_y^2(\tau)$ versus $\tau$, confidence intervals are negligible at left (short $\tau$) and widen moving to right (long $\tau$), where fewer terms are averaged. Confidence interval width is approximately proportional to $m^{-1/2}$, with $m$ equal to the number of averaged terms. In practice, being $N$ usually in the order of $10^4$ and above, $\text{Mod}\,\sigma_y^2(\tau)$ exhibits random ripple due to poor confidence only at the very right end of the curve.

## C. Equivalent Definition in the Frequency Domain

The MAVAR time-domain definition can be translated to an equivalent expression in the frequency domain, allowing a more profound understanding of the behaviour of this quantity. In fact, its definition can be rewritten as the mean square value of the signal output by a linear filter, with impulse response $h_{\text{MA}}(n, t)$ properly shaped according to definitions (7) and (8), receiving the signal $y(t)$, i.e.

$$\text{Mod}\,\sigma_y^2(\tau) = \left\langle \left[ \int_{-\infty}^{\infty} y(\xi)\, h_{\text{MA}}(n, t - \xi)\, d\xi \right]^2 \right\rangle = \left\langle \left[ y(t) * h_{\text{MA}}(n, t) \right]^2 \right\rangle \qquad (10).$$

The filter impulse response $h_{\text{MA}}(n, t)$ is plotted in Fig. 1 for example with $n=6$.

Hence, the MAVAR can be equivalently defined in the frequency domain as the area under the PSD of the signal output by this filter, i.e.

$$\text{Mod}\,\sigma_y^2(\tau) = \int_0^{\infty} S_y(f) \left| H_{\text{MA}}(n, f) \right|^2 df = \int_0^{\infty} S_y(f) \frac{2 \sin^6 \pi \tau f}{(n \pi \tau f)^2 \sin^2 \pi \dfrac{\tau}{n} f}\, df \qquad (11)$$

where $S_y(f)$ is the one-sided PSD of $y(t) = x'(t)$ and thus $S_y(f) = S_x(f) \cdot (2\pi f)^2$. The square magnitude of the filter transfer function $H_{\text{MA}}(n, f)$ is plotted in Fig. 2 for some values of the parameter $n$, having normalized the Fourier frequency $f$ to the inverse $1/\tau$ of the observation interval.

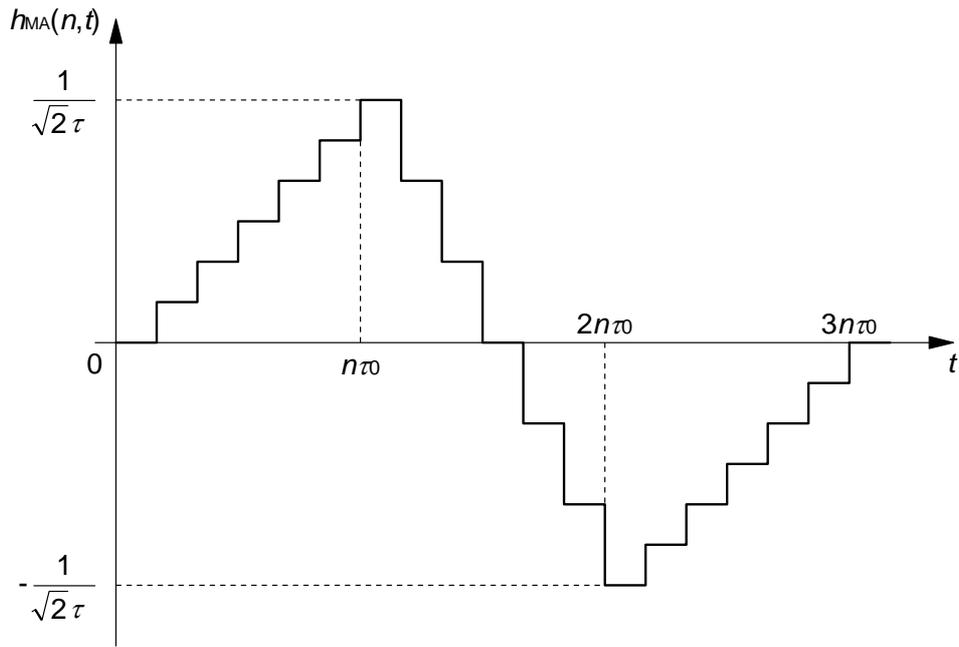

Fig. 1: Impulse response $h_{MA}(n,t)$ of the filter associated to the definition of Modified Allan Variance for $n$=6.

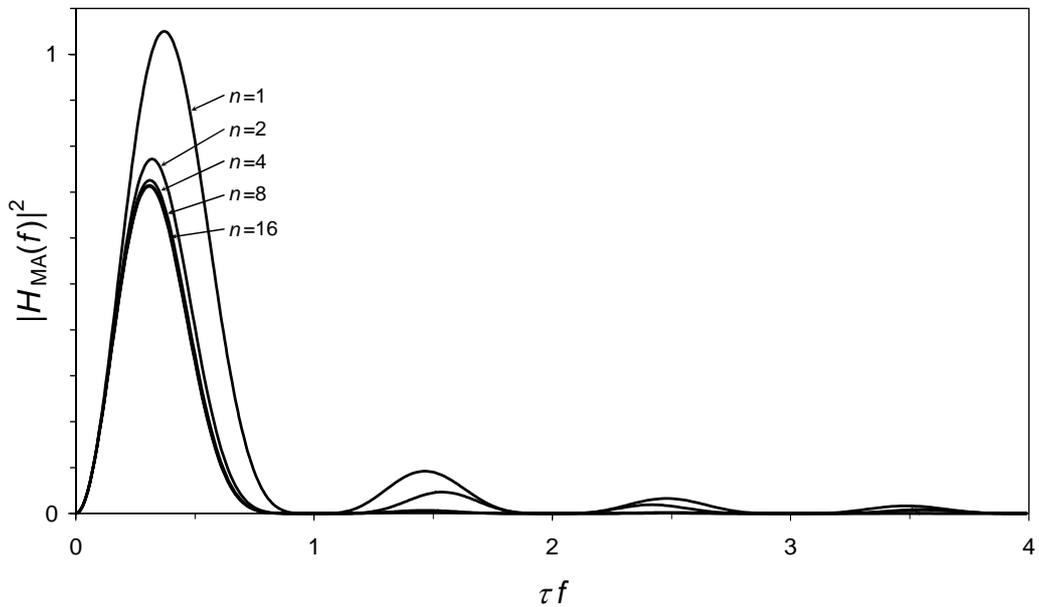

Fig. 2: Square magnitude of the transfer function $H_{MA}(n,t)$ associated to the definition of Modified Allan Variance for increasing values of the parameter $n$.

The limit transfer function for $n \to \infty$ and keeping $n\tau_0 = \tau$ constant is

$$\lim_{\substack{n \to \infty \\ n\tau_0 = \tau}} |H_{\mathrm{MA}}(n, f)|^2 = 2 \frac{\sin^6 \pi \tau f}{(\pi \tau f)^4} \tag{12}.$$

From Fig. 2, it can be seen that this limit is approached quickly for fairly low values of $n$ ($n > 4$).

Interesting enough is to notice that the transfer function $H_{\mathrm{MA}}(n, f)$ is pass-band, having magnitude shaped with a narrow main lobe centred at about $f \cong 1/(3\tau)$. Hence, $\mathrm{Mod}\,\sigma_y^2(\tau)$ gathers signal power selectively from a narrow band centred on a frequency inversely proportional to the observation interval $\tau$ (cf. eq. (11)). Thus, high-resolution spectral analysis of $y(t)$ can be achieved by computing $\mathrm{Mod}\,\sigma_y^2(\tau)$ over a range of $\tau$ values. References [13][27] treat this important topic in depth for various other variances defined for phase and frequency noise characterization.

### D. Behaviour with Power-Law Random Noise

It is convenient to generalize the LRD power-law model of spectral density (4). As customary in characterization of phase and frequency noise of precision oscillators [13], we will deal with random processes $x(t)$ whose one-sided PSD is modelled as

$$S_x(f) = \begin{cases} \displaystyle\sum_{i=1}^{P} h_{\alpha_i} f^{\alpha_i} & 0 < f \le f_{\mathrm{h}} \\ 0 & f > f_{\mathrm{h}} \end{cases} \tag{13}$$

where $P$ is the number of power-law noise types considered in the model, $\alpha_i$ and $h_{\alpha_i}$ are model parameters ($\alpha_i, h_{\alpha_i} \in \Re$) and $f_h$ is the unavoidable upper cut-off frequency.

As anticipated in Sec. III.A, power-law noise with $-4 \le \alpha_i \le 0$ has been revealed in practical measurements of various physical and natural phenomena, such as phase noise of precision oscillators and Internet traffic (see again note ), whereas $P$ should be not greater than few units for the model being useful. If the process $x(t)$ is simple LRD (4), then $P=1$ and $-1 < \alpha_i < 0$. Finally, the case $\alpha_i > 0$ is less interesting and will be not considered in this work [2].

Under this general hypothesis of power-law PSD, first we notice that, since $|H_{MA}(n,f)|^2$ in integral relationship (11) behaves as $\sim f^2$ for $f \rightarrow 0$, MAVAR convergence is ensured for $\alpha_i > -5$. Then, by considering separately each term of the sum in (13) and letting $P = 1$, $\alpha = \alpha_i$, evaluation of (11) with $S_y(f) = S_x(f) \cdot (2\pi f)^2$ yields corresponding time-domain expressions of $\text{Mod } \sigma_y^2(\tau)$. Complete formulas, for this and other variances, are reported in [6]; see also [10] for further details.

In summary, for any value in the whole range of convergence $-5 < \alpha \leq 0$ and under some other hypotheses commonly verified, MAVAR is found to obey asymptotically (ideally for $n \rightarrow \infty$, keeping constant $n\tau_0 = \tau$, but in practice for $n>4$, cf. Fig. 2) to a simple power law of the observation time $\tau$, i.e.

$$\text{Mod } \sigma_y^2(\tau) \sim A_\mu \tau^\mu \qquad (14)$$

where $\mu = -3 - \alpha$. If $P>1$, it is immediate to generalize (14) in summation of powers $A_{\mu_i} \tau^{\mu_i}$.

This is a fundamental result. If $x(t)$ obeys (13), a log-log plot of $\text{Mod } \sigma_y^2(\tau)$ looks ideally as a broken line made of $P$ straight segments, whose slopes $\mu_i$ can be measured to estimate the exponents $\alpha_i = -3 - \mu_i$ of the power-law noise components prevailing in distinct ranges of observation interval $\tau$.

### E. Behaviour with Deterministic Signals

It is important to understand the behaviour of MAVAR, as for any analysis tool of LRD and self-similar processes, also when $x(t)$ cannot be modelled simply as (13). In particular, in Internet traffic analysis, it is of outmost interest to know the behaviour of MAVAR when $x(t)$ includes deterministic components such as for instance offset from null mean, linear and quadratic drifts, periodic signals and steps, which are common and major examples of nonstationarity in data analyzed.

*1) Offset and drift.* Let us assume that $x(t)$ includes an offset mean value and a linear and quadratic drift, i.e. $x(t) = A + Bt + Ct^2$. By substitution in the MAVAR definition (6), we get

$$\text{Mod } \sigma_y^2(\tau) = 2C^2\tau^2 \tag{15}.$$

Thus, $\text{Mod } \sigma_y^2(\tau)$ is independent on data constant offset and linear drift, as obvious, but it reveals a quadratic drift in input data, resulting proportional to $\sim \tau^2$. Note that such a dependence on $\tau$ is the same as that yielded by power-law random noise with $\alpha \rightarrow -5$, which is found very seldom anyway.

*2) Periodic Signals.* Let us assume that $x(t)$ is a pure sine wave at frequency $f_m$, i.e. $x(t) = A \sin 2\pi f_m t$, with PSD $S_x(f) = (A^2/2)\cdot\delta(f - f_m)$. Then, by substitution in (11) and (12), we get (for $n \rightarrow \infty$, $n\tau_0 = \tau$)

$$\text{Mod } \sigma_y^2(\tau) = A^2 \frac{\sin^6 \pi f_m}{\left(\pi f_m\right)^4} \tag{16}.$$

Hence, the $\text{Mod } \sigma_y^2(\tau)$ plot exhibits a ripple with period $2/f_m$.

*3) Steps.* Major examples of nonstationarity in Internet traffic traces are sudden changes of the average bit rate, due for instance to traffic rerouting or link capacity adjustment. Let us assume that $x(t)$ is a step occurring at $t=0$, i.e. $x(t) = Au(t)$ (let $u(t)=0$ for $t<0$ and $u(t)=1$ for $t\geq0$). Since $y(t) = x'(t) = A\delta(t)$, from eq. (10) we get

$$\text{Mod } \sigma_y^2(\tau) = \left\langle \left[A \cdot h_{\text{MA}}(n,t)\right]^2 \right\rangle = 0 \tag{17}.$$

That is, single steps in the signal $x(t)$ ideally yield null impact on $\text{Mod } \sigma_y^2(\tau)$, defined by infinite-time averaging. In practical measurements, $x(t)$ has finite duration $T = (N-1)\tau_0$ and $\text{Mod } \sigma_y^2(\tau)$ is estimated using (9), thus being dependent on both $\tau$ and $T$. Nonetheless, the actual impact on $\text{Mod } \sigma_y^2(\tau, T)$ of input steps superposed to fractional noise is still negligible in most practical cases, provided that $T$ is long enough, as shown forth in Sec. V (Fig. 7).

## IV. ESTIMATING THE HURST PARAMETER OF LRD TIME SERIES BY MAVAR ANALYSIS

Let us consider a LRD process with PSD (4) characterized by Hurst parameter $1/2 < H < 1$. Then, from (5) and (14) and under all hypotheses made, $\text{Mod}\,\sigma_y^2(\tau)$ obeys a power law $\propto \tau^\mu$ (ideally for $n \to \infty$ but in practice for $n > 4$) with exponent

$$\mu = 2H - 4 \qquad (18).$$

In summary, it is possible to estimate the Hurst parameter $H$ of a sample realization $\{x_k\}$, supposed LRD with PSD (4), by the following procedure:

1) compute $\text{Mod}\,\sigma_y^2(\tau)$ with the estimator (9) based on the data sequence $\{x_k\}$ for increasing integer values $1 \le n < N/3$ (we always use a geometric progression of ratio 1.1, i.e. 24 values per decade, to ensure finest representation of the MAVAR trend);

2) estimate its average slope $\mu$ in a log-log plot for $n > 4$ (it is advisable to exclude also highest values of $n$, where confidence is lowest), by best fitting a straight line to the curve (e.g., by least square error; the actual fitting method does not impact significantly on the result);

3) check that $-3 < \mu < -2$, to confirm that $-1 < \alpha < 0$ ($0 < \gamma < 1$), and get the estimate of the Hurst parameter as $H = \mu/2 + 2$.

Under the more general hypothesis of power-law PSD (13), as noticed in the final remark of Sec. III.D, then up to $P$ slopes $\mu_i$ can be estimated ($-3 \le \mu_i < +2$) to yield the exponents $\alpha_i = -3 - \mu_i$ ($-5 < \alpha_i \le 0$) of the fractional noise or fractional Brownian motion components prevailing in distinct ranges of $\tau$.

Finally, some care should be exercised against the presence in data analyzed of deterministic components (e.g., steps), which cause trends in $\text{Mod } \sigma_y^2(\tau)$ that may be erroneously ascribed to random power-law noise. Sec. III.E already studied the behaviour of MAVAR when $x(t)$ includes some of the most common deterministic signals that yield nonstationarity in data analyzed.

## V. METHOD VALIDATION AND ACCURACY EVALUATION

The validity and accuracy of the method outlined in the previous section were evaluated by comparison to the well-established logscale diagram technique based on wavelet analysis [2][4]. All logscale diagrams were computed using standard scripts [28] (Daubechies' wavelet with three vanishing moments).

Both methods were applied to LRD pseudo-random data series $\{x_k\}$ of length $N$, generated with power-law one-sided PSD $S_x(f) = hf^\alpha$ $(-1 < \alpha \leq 0)$ for assigned values of $H = (1-\alpha)/2$. The generation algorithm is by Paxon [29]. In brief, it is based on spectral shaping: a vector of random complex samples, with mean amplitude equal to the square root of the desired value of $S_x(f_k)$ and phase uniformly distributed in $[0, 2\pi]$, is inversely Fourier-transformed to yield the time-domain sequence $\{x_k\}$. We evaluated MAVAR on several sample pseudo-random series so generated, producing $\text{Mod } \sigma_y^2(\tau)$ log-log plots almost linear and with expected slope, confirming the correctness of the generation procedure.

First, 10 pseudo-random sequences $\{x_k\}$ of length $N = 131072$, with mean 0 and variance 1, were generated for each of the 11 values $\{H_i\} = \{0.50, 0.55, 0.60, ..., 1.00\}$, corresponding to $\{\alpha_i\} = \{0, -0.1, -0.2, ..., -1.0\}$. On the resulting 110 time series, we applied both the MAVAR and the LD methods, getting two sets of estimates of $H$, respectively $\{\hat{H}_{i,j}^{\text{MA}}\}$ and $\{\hat{H}_{i,j}^{\text{LD}}\}$, for $i = 0, 1, ..., 10$ and $j = 1, 2, ..., 10$. We then evaluated the accuracy of these estimates with respect to the assigned generation values $H_i$, calculating the estimation errors $\Delta_{i,j}^{\text{MA}} = \hat{H}_{i,j}^{\text{MA}} - H_i$ and $\Delta_{i,j}^{\text{LD}} = \hat{H}_{i,j}^{\text{LD}} - H_i$.

Furthermore, we decided to compare the accuracy attained by the two methods on short sequences, when results are impaired by poor confidence. Thus, we repeated the same test as before, but on other two sets of 110 sequences of length $N = 1024$ and $N = 2048$, respectively.

Fig. 3 compares the absolute estimation errors $\left\{ \Delta_{i,j}^{\mathrm{MA}} \right\}$ and $\left\{ \Delta_{i,j}^{\mathrm{LD}} \right\}$ attained by the two methods on sequences of $N = 131072$ samples. For each value $H_i$, the mean and standard deviation of the 10 estimation errors are plotted as dots with $\pm\sigma$ vertical bars. Whereas both methods feature excellent accuracy on such a long data sequence, limiting estimation errors within $\pm1\%$ or $\pm2\%$ of the target value $H_i$, yet we notice that MAVAR results exhibit better confidence than those of LD, because standard deviation of estimates is much smaller.

Similarly, Figs. 4 and 5 compare the absolute estimation errors $\left\{ \Delta_{i,j}^{\mathrm{MA}} \right\}$ and $\left\{ \Delta_{i,j}^{\mathrm{LD}} \right\}$ attained by the two methods on short sequences of $N = 2048$ and $N = 1024$ samples, respectively. Here, the better accuracy and confidence of MAVAR results is even more evident: not only the standard deviation of MAVAR estimates is much smaller than that of LD, but also the mean of LD estimates departs significantly from the target value $H_i$ in most cases. This mean error offset from zero was noticed also in other simulation results presented in [30], for $N = 1000$ and $N = 50000$.

Visual comparison of MAVAR and LD plots, not shown here for lack of space, justifies the better confidence of $H$ estimates achieved by MAVAR. Especially on short sequences ($N = 2048$ and $N = 1024$), MAVAR log-log plots are far smoother, closer to the ideal linear trend even at the right side, where confidence is worse.

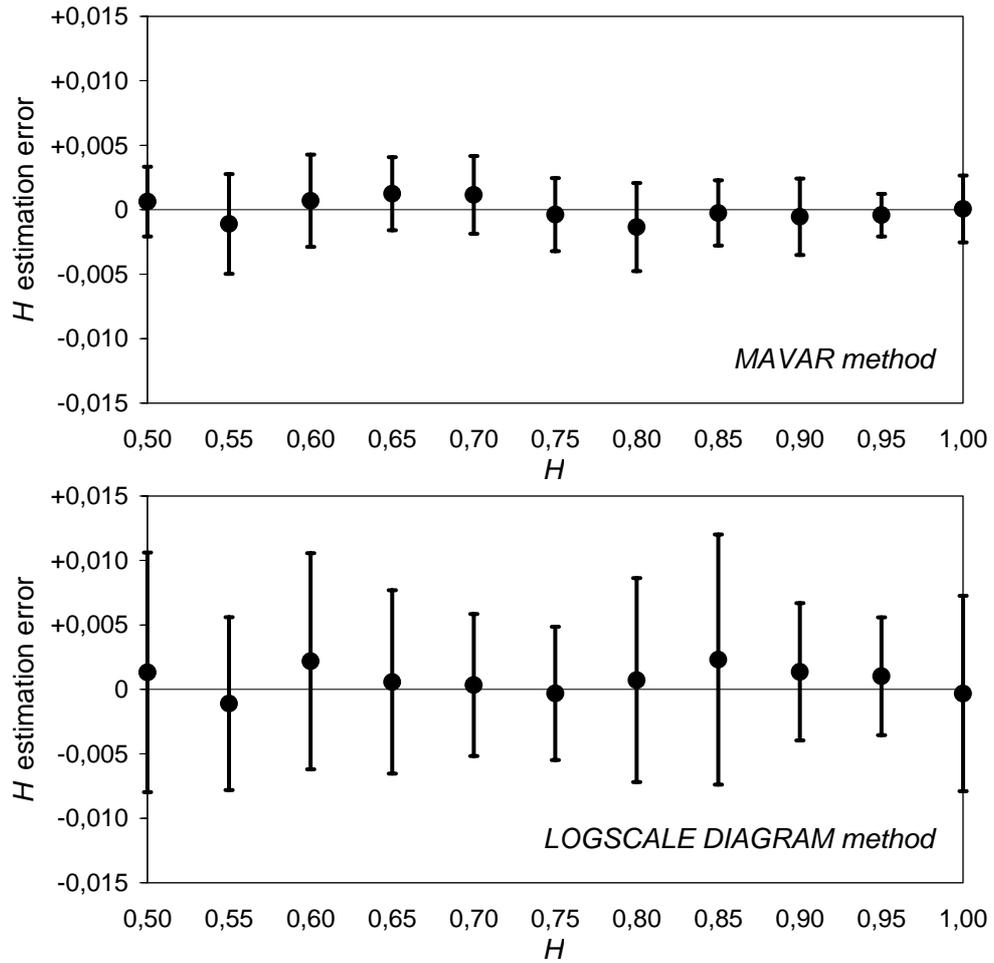

Fig. 3: Absolute estimation error of *H* attained by the MAVAR and LOGSCALE DIAGRAM methods (*N*=131072). For each value $H_i$ the mean and standard deviation of the 10 estimation errors are plotted.

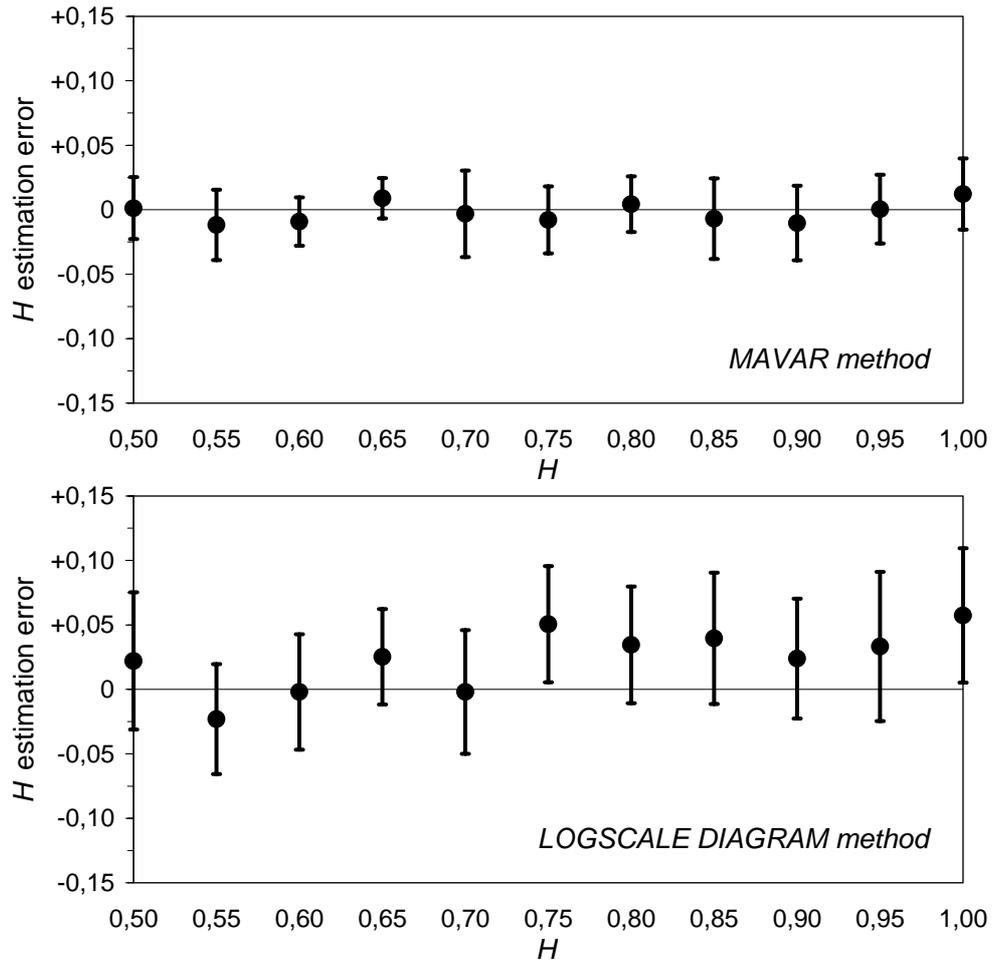

Fig. 4: Absolute estimation error of *H* attained by the MAVAR and LOGSCALE DIAGRAM methods (*N*=2048). For each value $H_i$ the mean and standard deviation of the 10 estimation errors are plotted.

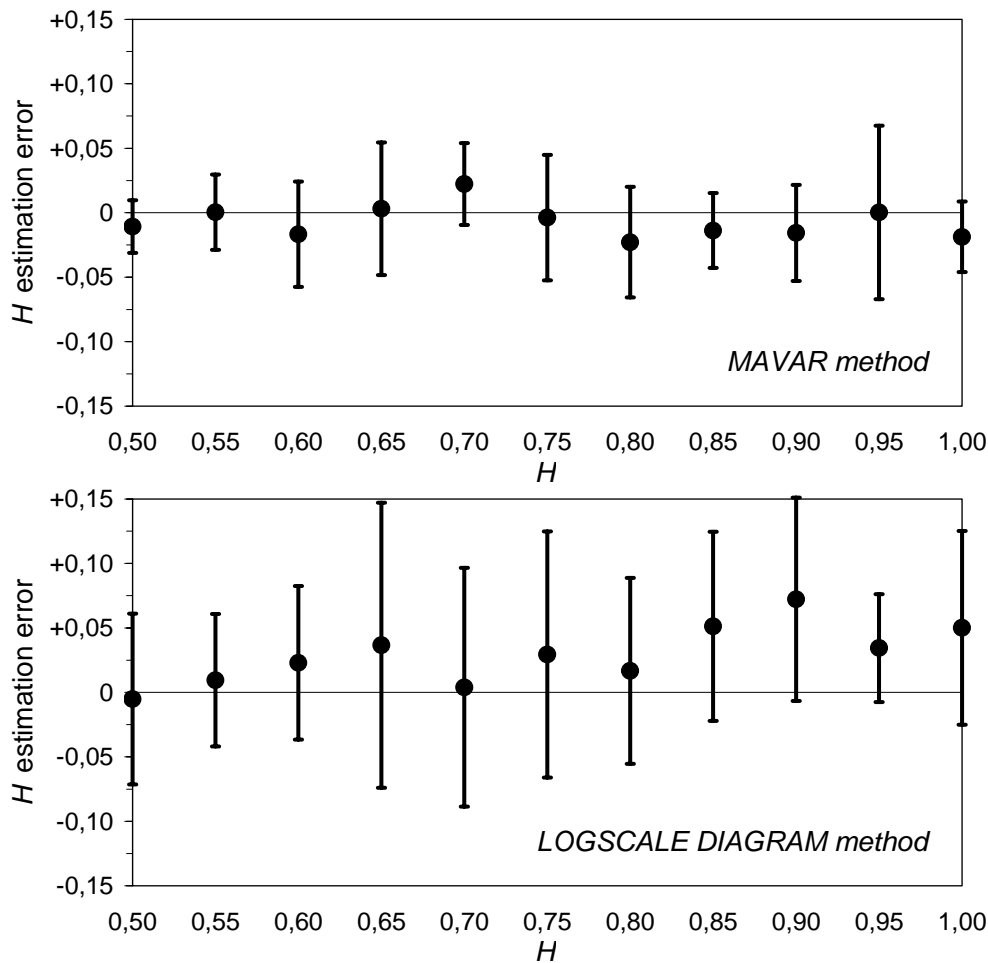

Fig. 5: Absolute estimation error of *H* attained by the MAVAR and LOGSCALE DIAGRAM methods (*N*=1024). For each value $H_j$ the mean and standard deviation of the 10 estimation errors are plotted.

To highlight the impact of the sample sequence length *N* on the uncertainty of the estimate of *H*, we applied the MAVAR method to four different pseudo-random sequences generated with *H*=0.75, truncating their length subsequently to increasing values $1000 \leq N \leq 50000$. The graph in Fig. 6 plots the resulting estimation error as *N* increases.

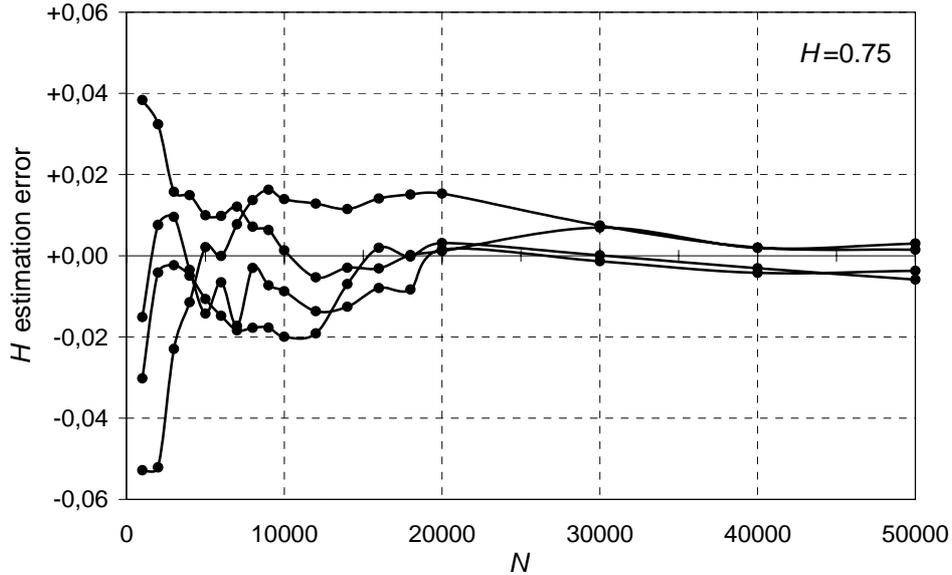

Fig. 6: Convergence of the MAVAR method in estimating $H$=0.75 of four different pseudo-random sequences with truncated length $1000 \leq N \leq 50000$.

Finally, we evaluated MAVAR and logscale diagrams on input data with steps of various amplitude superposed to LRD noise, as discussed in Sec. III.E.3. To this aim, input sequences of length $N = 1024$ and $N = 131072$ were generated as $\{x_k\} = \{Au_{k\text{-}M} + n_k\}$ ($k = 1, 2, …, N$), where $\{u_{k\text{-}M}\}$ is the sampled version of the unit step function $u(t)$ delayed $M$ time units ($1 < M < N$) and $\{n_k\}$ is a pseudo-random LRD series, with mean $m_n$=0 and variance $\sigma_n^2$=1, generated as before with power-law one-sided PSD $S_n(f) = hf^\alpha$ for $\alpha = $ -0.60 ($H = 0.80$). By varying extensively parameters $M$ and $A$, we found that:

❑ the impact of the step on $\text{Mod}\,\sigma_y^2(\tau)$ is maximum for $M \cong N/2$ (i.e., when the step occurs at half the sequence), whereas it tends to be negligible for $M$ approaching the extremes 1 and $N$;

❑ the impact of the step on $\text{Mod}\,\sigma_y^2(\tau)$ is noticeable for $N = 1024$, but very little for $N = 131072$ (cf. eq. (17));

❑ even for $N = 1024$, the step amplitude $A$ must be at least on the order of the LRD noise standard deviation (1 in this case) to impact significantly $\text{Mod}\,\sigma_y^2(\tau)$; however, such a big step is evident by simple visual inspection in input data and thus can be easily identified and removed before MAVAR computation, to avoid erroneous identification of random power-law noise;

❑ in most cases, steps in input data affect $\mathrm{Mod}\,\sigma_y^2(\tau)$ less than LD.

Among the numerous simulation results collected, Fig. 7 show subsets of curves obtained on the LRD sequence of length $N = 131072$, varying step amplitude $A = 0$, 0.5, 1, 2 and delay $M = 0.05N$, $0.25N$, $0.50N$, $0.75N$, $0.95N$. In comparing graphs, it should be considered that $\mathrm{Mod}\,\sigma_y^2(\tau)$ is plotted over almost the full range $1 \leq n \leq N/3$, whereas LD omits the last scales $j > 14$, where confidence is scarce or null and step impact is more evident. In the MAVAR graph, moreover, we notice that step impact is little and mostly limited to the right end, where curve slope should not be considered anyhow due to poor confidence.

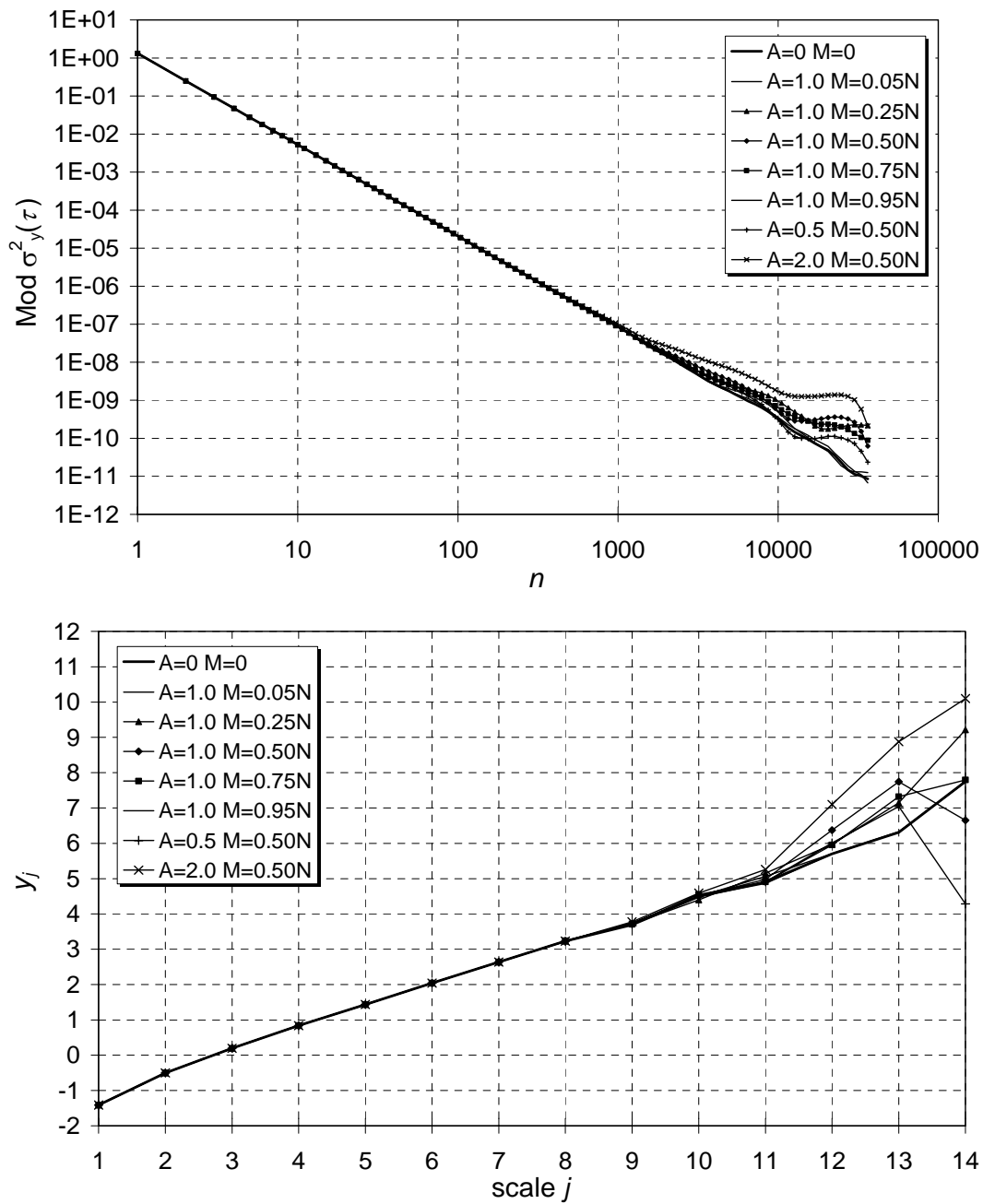

Fig. 7: MAVAR (upper graph) and LD (lower graph) computed on a pseudo-random LRD sequence $\{n_k\}$ of length $N$=131072 ($m_n$=0, $\sigma_n^2$=1, $H$ = 0.80) with added step $\{Au_{k-M}\}$ for different values of amplitude $A$ and delay $M$.

## VI. Application to a Real IP Traffic Trace

We tested the MAVAR and LD methods on a real IP traffic trace, obtained by counting the packets transmitted per time unit on a transoceanic system (MAWI Project [31]). The data sequence under analysis is made of $N=2^{16}=65536$ samples, acquired with sampling period $\tau_0=8$ ms, thus spanning a measurement interval $T\cong524$ s. No evident nonstationary trends, such as steps, are present.

Fig. 8 shows the logscale diagram, computed using standard scripts [28] (Daubechies' wavelet with three vanishing moments), with 95%-confidence intervals depicted as vertical bars. The slope of the straight line fitting a left portion of the curve (interval of scales $j=2$ to 10) yields the estimate $H\cong0.59$ ($\alpha\cong-0.18$). We notice that, because of the irregular trend of the curve, this estimate depends significantly on the interval on which the average slope is calculated. The precise choice of the interval 2÷10 is arbitrary and thus this $H$ estimate results uncertain.

Moreover, at the right side of the curve, an average slope increase is evident, revealing a different scaling behaviour at higher scales. The multislope trend and the ample confidence bars make any parameter estimation questionable in this area. However, we note that the average end-to-end slope of the curve on scale interval 10÷14 is $\alpha\cong1.25$.

On the other hand, Fig. 9 shows the Modified Allan Variance computed on the same trace for 10 ms $<\tau<170$ s (as before, about 24 points per decade). Compared to LD, MAVAR gives a clearer picture of the spectral characteristics of the traffic sequence under analysis. The MAVAR curve, in log-log scale, is very well approximated by two linear segments, whose slopes are $\mu_1=-2.824$ (10 ms $<\tau<10$ s) and $\mu_2=-1.80$ (10 s $<\tau<100$ s). Almost no spurious ripples are visible in those intervals, in spite of the high density of points in which MAVAR has been computed.

Therefore, according to the power-law model (13), two simple components are revealed by MAVAR: a main one with $\alpha_1=-0.176$, dominant over three decades of observation time for 10 ms $<\tau<10$ s, and a secondary one with $\alpha_2=-1.20$, dominant for 10 s $<\tau<100$ s. The former noise is LRD-type, with $H\cong0.59$. These estimates, for both noise components, are in agreement with LD results. However, analysis and simulation results reported in this work ensure that MAVAR estimates are more accurate and with better confidence.

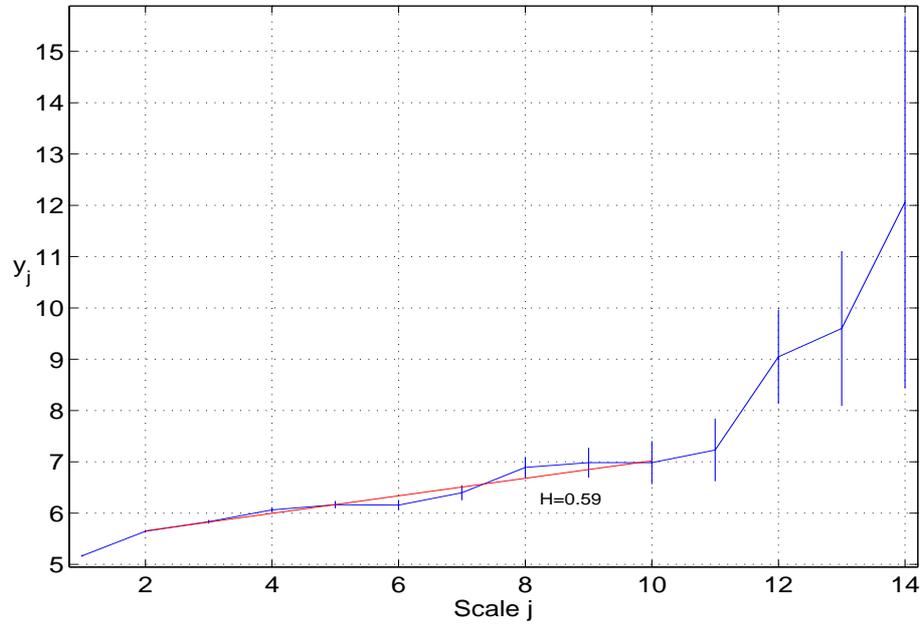

Fig. 8: Logscale diagram of a real IP packet/time trace
(MAWI Project [31], $N$=65536, $\tau_0$=8 ms, $T \cong$524 s).

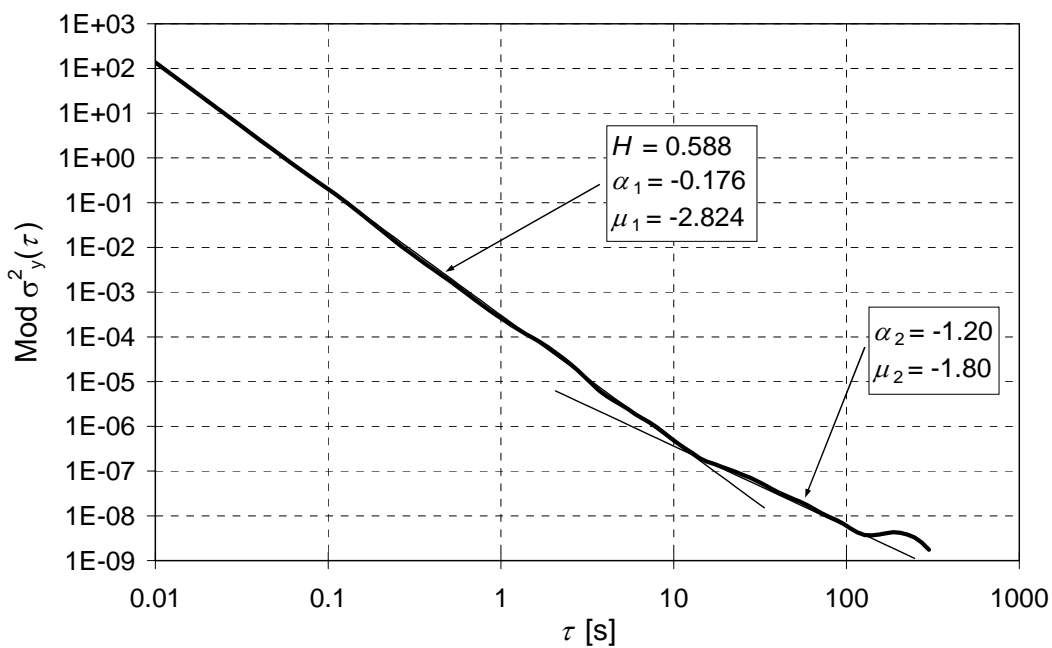

Fig. 9: Modified Allan Variance of a real IP packet/time trace
(MAWI Project [31], $N$=65536, $\tau_0$=8 ms, $T\cong$524 s).

## VII. CONCLUSIONS

In this paper, we proposed to use the Modified Allan Variance, a time-domain tool originally conceived for frequency stability characterization, for estimating the Hurst parameter of LRD traffic traces and, more generally, for identifying fractional noise components in network traffic. MAVAR definition and properties more relevant to this aim were summarized. Moreover, the behaviour of MAVAR with most common deterministic signals that yield nonstationarity in data under analysis was also studied.

The MAVAR method was validated by comparison to one of the best techniques for analyzing self-similar and LRD traffic: the logscale diagram based on wavelet analysis. Both methods were tested on pseudo-random LRD data series, generated with assigned values of $H$. MAVAR proved very accurate and exhibited good confidence in estimating $H$, even better than the LD method, despite its computational lightweight. Moreover, both methods were tested on sequences where a step is superposed to LRD noise, simulating one of the major examples of nonstationarity in Internet traffic.

Estimation of LRD noise parameters ($\gamma$, $H$) by MAVAR proved quite robust against steps in input data; the actual impact of steps on $\text{Mod}\,\sigma_y^2(\tau)$ is limited or negligible in most practical cases.

Finally, both methods were applied on a real IP traffic trace, providing a sound example of the usefulness of MAVAR also in traffic characterization. While the LD produced uncertain $H$ estimate, due to the irregular trend of the curve, MAVAR gave a clearer picture of the spectral characteristics of the traffic sequence under analysis. Two simple power-law noise components were revealed, with PSD $k_1/f^{0.176} + k_2/f^{1.20}$. The first term, dominant over three decades of observation time for $10\text{ ms} < \tau < 10\text{ s}$, is LRD with $H \cong 0.59$.

As a final remark, we point out that MAVAR is not proposed as ultimate or best tool for traffic analysis. Rather, we believe that it may complement usefully other established techniques, such as the LD method, due to its several advantages. Among them, we highlight:

❑ high spectral sensitivity (cf. Fig. 2);

❑ convergence to finite values for all power-law noise types with $\alpha > -5$, which implies direct applicability to fractional noise or fractional Brownian motion noise without specific pre-processing (differentiation);

❑ ability to estimate accurately parameters of LRD and generic power-law processes ($\gamma$, $H$, $\alpha_i$) over the full range of convergence (cf. Figs. 3, 4 and 5);

❑ efficient use of input data, yielding good confidence of parameter estimates (cf. again Figs. 3, 4 and 5);

❑ possibility to evaluate $\text{Mod}\,\sigma_y^2(n\tau_0)$ for any integer value of $n$ (even all values $1 \le n < N/3$), allowing finest representation of the MAVAR curve and thus optimal identification of MAVAR trends;

❑ robustness against various common nonstationary components in data analyzed.

## BIOGRAPHY OF STEFANO BREGNI

**Stefano Bregni** (M'92-SM'99) was born in Milano, Italy, in 1965. He received his MSc degree in Telecommunications Engineering from Politecnico di Milano. In 1991, he joined SIRTI S.p.A., where he was involved in SDH transmission systems testing and in network synchronization issues, with special regard to clock stability measurement. From 1994 to 1999, he was with CEFRIEL (consortium of private companies with Politecnico di Milano) as head of the Transmission Systems Dept. He is Associate Professor at Politecnico di Milano (tenured Asst. Prof. since 1999), where he teaches telecommunications networks and transmission networks.

He has been Senior Member of IEEE since 1999. He is author of the book *"Synchronization of Digital Telecommunications Networks"*, John Wiley & Sons, 2002, of another book on SDH systems, McGraw Hill Italia, 2004, and of about fifty papers on various telecommunications topics. He is Distinguished Lecturer of the IEEE Communications Society (Expert Lecturer since 1999). He was vice-chair of the Transmission, Access and Optical Systems Technical Committee of the IEEE Communications Society from 1999 to 2003, retaining the position of secretary from 2004. He has been appointed co-chair or vice-chair of symposia in IEEE ICC 2004, GLOBECOM 2005 and ICC2006 conferences. He served in the Technical Program Committees of several other ICC and GLOBECOM conferences.

## BIOGRAPHY OF LUCA PRIMERANO

To be added.